\documentclass[english,keywords,aps,amsmath,amssymb,twocolumn]{revtex4-1}
\usepackage[T1]{fontenc}
\usepackage[latin1]{inputenc}
\usepackage{babel}
\usepackage{amssymb}
\usepackage{graphicx}
\usepackage{color}
\usepackage{bm}
\usepackage{longtable}
\usepackage{amsmath} 
\usepackage{amsfonts}
\usepackage{amssymb}
\begin{document}

\title{CHSH inequalities with appropriate response function for POVM and their quantum violation}
\author{Asmita Kumari}
\author{A. K. Pan \footnote{akp@nitp.ac.in}}
\affiliation{National Institute Technology Patna, Ashok Rajpath, Patna, Bihar 800005, India}
\begin{abstract}
In the derivation of local bound of a Bell's inequality,  the response functions corresponding to the different outcomes of measurements are fixed by the relevant hidden variables irrespective of the fact if the measurement is unsharp. In the context of a recent result by Spekkens \cite{spekkens2014} that tells even in an ontological model the unsharp observable cannot be assigned a deterministic response function, we derive a modified local bound of CHSH inequality in unsharp measurement scenario. We consider response function for a given POVM which is determined by the response functions of the relevant projectors appearing in its spectral representation. In this scenario, the local bound of CHSH inequality is found to be dependent on the unsharpness parameter. This then enables us to show that the quantum violation of CHSH inequality for unbiased spin-POVMs occurs whenever there is violation for their sharp counterpart. For the case of biased POVMs, it is shown that the quantum violation of CHSH inequality can be obtained for ranges of sharpness parameter for which no violation obtained using standard local bound of CHSH inequality.

\end{abstract}
\maketitle
\section{Introduction}
Bell's theorem is considered to be one of the most fundamental result in the foundations of quantum mechanics (QM). It states that the attempts of  providing more complete specification of reality than quantum theory in terms of a local ontological model is not possible \cite{bell64,chsh69,eberhard93}. In a local model, it is assumed that a suitable set of ontic states ($\lambda$s) assign the response functions of different measurement outcomes (the assumption of reality) and such response functions remain uninfluenced by space-like separated measurements and their outcomes (the assumption of locality). Based on this assumption of local realism (LR) in an ontological model, Bell's inequalities is derived which is shown to be violated by QM. Clauser, Horne, Shimony and Holt (CHSH) is the simplest and commonly referred Bell's inequalities defined in two-party, two-measurement and two-outcome scenario \cite{chsh69}. 

Before proceeding further let us recapitulate the essence of an ontological model reproducing the quantum statistics \cite{harrigan10}. Given a preparation procedure $P\in \mathcal{P}$ and a measurement procedures $M\in \mathcal{M}$, an operational theory assigns probability $p(k|P, M)$ of obtaining a particular outcome $k\in \mathcal{K}_{M}$. Here $\mathcal{M}$ is the set of measurement procedures and $\mathcal{P}$ is the set of preparation procedures. In QM, a preparation procedure produces a density matrix $\rho$ and measurement procedure, in general, is described by a suitable POVM $E_k$ and finally the theory provides a rule for the probability of a particular outcome $ k $, given by $p(k|P, M)=Tr[\rho E_{k}]$, which is known as Born rule. In an ontological model of QM, it is assumed that whenever $\rho$ is prepared by a specific preparation procedure $P\in \mathcal{P}$ a probability distribution $\mu_{P}(\lambda|\rho)$ in the ontic space is prepared, satisfying $\int _\Lambda \mu_{P}(\lambda|\rho)d\lambda=1$ where $\lambda \in \Lambda$ and $\Lambda$ is the ontic state space. The probability of obtaining an outcome $k$ is given by a response function $\xi_{M}(k|\lambda, E_{k}) $ satisfying $\sum_{k}\xi_{M}(k|\lambda, E_{k})=1$ where a measurement operator $E_{k}$ is realized through a particular measurement procedure $M\in\mathcal{M}$. A viable ontological model should reproduce the Born rule, i.e., $\forall \rho $, $\forall E_{k}$ and $\forall k$, $\int _\Lambda \mu_{P}(\lambda|\rho) \xi_{M}(k|\lambda, E_{k}) d\lambda =Tr[\rho E_{k}]$.

Let two parties Alice and Bob performs measurements of two dichotomic observables  $(A_{1}, A_{2})$ and $(B_{1}, B_{2})$ respectively at space-like separated sites. The well-known CHSH inequalities provide the local bound of a suitable algebraic combinations of the correlations $\langle A_i B_j\rangle$ with $i,j =1,2$.  One of the four CHSH expressions can be written as 
\begin{eqnarray}
\label{bell1}
\left\langle \mathbb{B}\right\rangle  = \left\langle A_1 B_1\right\rangle  +\left\langle A_1 B_2\right\rangle +\left\langle A_2 B_1\right\rangle -\left\langle A_2 B_2\right\rangle 
\end{eqnarray}

In an ontological model, the correlation $\langle A_i B_j\rangle$ can be calculated as 
\begin{eqnarray}
\nonumber
\langle A_i B_j\rangle_{\lambda}=\sum_{a_i,b_j=\pm 1}a_i b_j\int_{\Lambda}\mu(\lambda|\rho)\xi( a_i b_j|\Pi^{a_i}_{A_i},\Pi^{b_j}_{B_j},\lambda)  d\lambda
\end{eqnarray}
where $\Pi^{a_i}_{A_i}$ and $\Pi^{b_j}_{B_j}$ are the measurement operators corresponding to the observables $A_i$ and $B_j$ respectively. Unlike the Kochen-Specker (KS) scenario \cite{ks,spekk05}, where $\lambda$ predicts the outcomes deterministically, in Bell scenario, the response function fixed by $\lambda$ can  be probabilistic. The crucial point in Bell scenario is the condition of factorizability - the conditional independence of the response function of space-like separated measurements. Mathematically, $\xi( a_i b_j|\Pi^{a_i}_{A_i},\Pi^{b_j}_{B_j},\lambda) = \xi( a_i|\Pi^{a_i}_{A_i},\lambda) \xi( b_j|\Pi^{b_j}_{B_j},\lambda)$. Note that factorizability is a weaker constraint than
outcome determinism. The latter implies the former but the converse is not true in general. Using factorizability condition, we can write
\begin{eqnarray}
\label{ab}
\langle A_i B_j\rangle_{\lambda} &=& \int _\Lambda  \mu(\lambda|\rho) \overline{p}( a_i |\Pi^{a_i}_{A_i},\lambda) \overline{p}( b_j|\Pi^{b_j}_{B_j},\lambda)  d\lambda 
\end{eqnarray}
where $\overline{p}( a_i |\Pi^{a_i}_{A_i},\lambda)=\sum_{a_i} a_i \xi( a_i|\Pi^{a_i}_{A_i},\lambda) $ and $\overline{p}( b_j|\Pi^{b_j}_{B_j},\lambda)=\sum_{b_i} b_i \xi( b_i|\Pi^{b_j}_{B_j},\lambda)$ such that $ \xi( a_i|\Pi^{a_i}_{A_i},\lambda), \xi( b_i|\Pi^{b_j}_{B_j},\lambda)  \in  \left\{0,1\right\}$. 

Using the correlation in an ontological model satisfying factorizability as defined in Eq.(\ref{ab}), we have the celebrated CHSH inequality is given by
\begin{eqnarray}
\label{Slocal}
\langle\mathbb{B}\rangle^S_{\lambda} = \int _\Lambda \mu(\lambda|\rho) \mathbb{B}^S_\lambda d\lambda \leq 2
\end{eqnarray}
By the superscript $`S'$ we mean the standard way of deriving of local bound of CHSH inequality. In QM, $\langle\mathbb{B}\rangle^S_{Q} \leq 2 \sqrt{2}$ thereby violating the CHSH inequality in Eq.(\ref{Slocal}).\\
Note that, in the standard derivation of the local bound given by Eq.(\ref{Slocal}) the values of the response functions $\xi(a_i|\Pi^{a_i}_{A_i},\lambda)$ and $\xi(b_j|\Pi^{b_j}_{B_j},\lambda)$ are assigned irrespective of the fact weather the concerned measurement operator, say $\Pi^{a_i}_{A_i}$, is a POVM or a projector. It is proved \cite{spekkens2014,hari07} that if a  POVM can be written as a convex combination of projectors then in an ontological model, the response function of that POVM will follow the same functional relation. Even, it is shown \cite{spekkens2014} that the response function of a POVM has to be necessarily indeterministic. Against this backdrop, let us first closely examine the Bell correlation scenario in an \textit{outcome deterministic} model, where $ \xi( a_i|\Pi^{a_i}_{A_i},\lambda), \xi( b_i|\Pi^{b_j}_{B_j},\lambda)  \in  \left[0,1\right]$. Does $\lambda$ also fix the response function of a POVM in such a model? 

 In order to find the answer to this question, consider  an unbiased spin POVM which  can be written as $E_{k\pm}=\frac{1\mp\eta}{2} \mathbb{I}\pm \eta P_{k}$ where $P_{k}$ is the projector. Then, in an ontological model one can write $\xi(k|E_{k+},\lambda)=\frac{1 - \eta}{2} + \eta \xi(k|P_{k},\lambda) $. If one assumes that $\lambda$ fixes the response function of the POVM deterministically, i.e.,  $\xi(k|E_{k+},\lambda)=1$, then it requires both $\eta=1$ and $\xi(k|P_{k},\lambda)=1$. The effect of unsharpness thus becomes redundant. On the other hand if $\lambda$ fixes the response function of projectors, i.e., $\xi(k|P_{k},\lambda)=1$ then, $\xi(k|E_{k\pm},\lambda)=\frac{1\pm\eta}{2}$. This simple example shows that in a outcome deterministic ontological model it is only  meaningful to assign the deterministic response function to the projectors. Note that, KS non-contextuality assumes outcome determinism for projectors along with the assumption of measurement non-contextuality. If CHSH inequalities are meant to test non-contextuality then such a deterministic model mentioned above fits perfectly with the spirit of KS non-contextuality. 

However, CHSH inequalities holds even for indeterministic response functions.  It is then curious whether the revised notion of fixing response function for POVM is valid for the case when $\lambda$ assigns the indeterministic response function to projectors. There is as such no logical reason why this cannot be done. In fact it is proved in \cite{spekkens2014}. In this work, we adopt the revised notion for assigning the response function for POVM  in order to derive the local bound of a CHSH inequality for both biased and unbiased POVMs. Such a conceptual update provides a modified local bound of CHSH inequality for unsharp measurement scenario. In literature, there are papers \cite{kar,busch} attempted to see Bell violation in the context of unsharp
measurement. We derived a different local bound of CHSH inequality which is function of sharpness parameters of spin-POVMs and smaller than the standard bound $2$. For the case of biased POVMs, the bound depends on both the sharpness and biasedness parameters. For the case of spin-POVM, we show that quantum violation of CHSH inequality can be achieved for any given state and measurement settings for which violation can be shown for sharp measurement and for the biased POVM case we show that the violation can be obtained for a range of values of the sharpness parameter when the violation cannot be obtained for the standard case.

\section{Local bound of CHSH inequality for unsharp measurements}
With the revised notion of fixing the response function of  a POVM in an ontological model, let us now derive the local bound of CHSH inequality for unsharp measurement scenario.
\subsection{For unbiased POVM}
Consider Alice and Bob performs the measurements of the POVMs $\Pi^{\pm}_{A_i}$ and $\Pi^{\pm}_{B_j}$ are respectively given by 
\begin{subequations}	
\begin{align}
\label{upovm}
\Pi^{\pm}_{A_i}&=\frac{\mathbb{I} \pm \eta   A_{i}}{2}\\
 \Pi^{\pm}_{B_j}&=\frac{\mathbb{I} \pm  \eta B_{j}}{2}
\end{align}
\end{subequations}	
where $ \eta $ $(0 \leq \eta \leq 1)$ is sharpness parameter. Since $A_{i}=\vec{a}_{i}.\sigma= (P^{+}_{A_i}-P^{-}_{A_i})$ and $B_j=\vec{b}_{j}.\sigma=(P^{+}_{B_j}-P^{-}_{B_j})$ then Eq.(\ref{upovm}) can be re-written as explicit convex combination of projectors is given by 
\begin{eqnarray}
\label{pv1}
{\Pi}^{\pm}_{A_i} = (\frac{1 \pm \eta}{2} P^{+}_{A_i}  + \frac{1 \mp \eta}{2} P^{-}_{A_i} )
\end{eqnarray}
and
\begin{eqnarray}
\label{pv2}
{\Pi}^{\pm}_{B_j} = (\frac{1 \pm \eta}{2}P^{+}_{B_j}  + \frac{1\mp \eta}{2} P^{-}_{B_j})
\end{eqnarray}
Since coarse graining of the measurement outcome is represented by coarse graining of the corresponding response function \cite{spekkens2014}, the response function assigned by the ontic state $\lambda$ for unbiased POVMs will be the convex combination of the response functions of projectors. Using this notion of fixing response function in an ontological model leads a different local bound of CHSH inequalities is given by
\begin{eqnarray}
\label{Usb}
\langle\mathbb{B}\rangle^{\eta}_{\lambda} = \int _\Lambda \mu(\lambda|\rho) \mathbb{B}^{\eta}_\lambda d\lambda \leq 2 \eta^2
\end{eqnarray}
which is explicitly dependent on sharpness parameters $\eta$, and lower than the standard local bound $\langle\mathbb{B}\rangle^S_{\lambda} \leq 2$. Clearly, $\langle\mathbb{B}\rangle^{\eta}_{\lambda} \leq \langle\mathbb{B}\rangle^S_{\lambda}$. The explicit derivation of the bound given by Eq.($\ref{Usb}$) is placed in Appendix A.\\

In QM, if Alice and Bob shares an entangled state
\begin{eqnarray}
\label{wer}
 |\psi \rangle = \cos(\theta)|0 0\rangle+ \sin(\theta)|1 1 \rangle
\end{eqnarray}
 Then, for the measurement directions $\vec{a}_{1}=  \hat{z}$, $\vec{a}_{2}= \hat{x}$, $\vec{b}_{1}=\frac{ \hat{z}+ \hat{x}}{\sqrt{2}}$ and $ \vec{b}_{2}=\frac{ \hat{z}- \hat{x}}{\sqrt{2}}$, the quantum bound of CHSH inequality is 
\begin{eqnarray}
\label{ubell}
\langle\mathbb{B}\rangle^{\eta}_{Q} \leq \sqrt{2} \eta^2   (1+\sin (2 \theta ))
\end{eqnarray}
Now standard CHSH inequality is violated if
\begin{eqnarray}
\label{s}
\Delta^{S}&=&\langle\mathbb{B}\rangle^{\eta}_{Q}-\langle\mathbb{B}\rangle^{S}_{\lambda}
\\  \nonumber 
&=& \sqrt{2} \eta^2   (1+\sin (2 \theta ))-2>0
\end{eqnarray}
It is seen from Eq.($\ref{s}$) that, the violation of standard CHSH inequality will be restricted by sharpness parameter $\eta$ along with $\theta$.

However, the difference of quantum bound and our revised local bound for unbiased spin-POVM is given by 
 \begin{eqnarray}
 \label{sdel}
 \Delta^{\eta}&=&\langle\mathbb{B}\rangle^{\eta}_{Q}-\langle\mathbb{B}\rangle^{\eta}_{\lambda}
 \\  \nonumber 
 &=& \sqrt{2} \eta^2   (1+\sin (2 \theta ))-2\eta^2
  \\  \nonumber 
  &=&  \eta^2(\sqrt{2}   (1+\sin (2 \theta ))-2),
 \end{eqnarray}
 For the violation of CHSH, $\Delta^{\eta}>0$ is required. Since $\eta$ is positive quatity it is enough to obtain violation if $\sqrt{2}   (1+\sin (2 \theta ))-2>0$ is satisfied which is the condition for sharp measurement. Hence, for the case of unbiased spin-POVM, if there is violation of standard CHSH inequality for sharp measurement for certain measurement settings, the same violation can be obtained for unsharp spin-POVM. One can easily show that the result extends to mixed state as well.    

\subsection{For biased POVM}
Let the POVMs of Alice and Bob for biased measurements are given by 
\begin{subequations}	
\begin{align}
\Pi^{\pm}_{A_i}&=\frac{\mathbb{I} \pm(\mathbb{I} \alpha +\eta   A_i)}{2}\\
 \Pi^{\pm}_{B_j}&=\frac{\mathbb{I} \pm (\mathbb{I} \alpha + \eta B_j)}{2}
\end{align}
\end{subequations}
 respectively, where $\alpha$ is the biasedness parameter such that $|\alpha|+\eta\leq1$. The POVMs $\Pi^{\pm}_{A_i}$ and $\Pi^{\pm}_{B_j}$ can be re-written as
\begin{align}
{\Pi}^{\pm}_{A_i}=    (\frac{1 \pm \alpha \pm \eta}{2} P^{+}_{A_i}  + \frac{1 \pm \alpha \mp \eta}{2} P^{-}_{A_i} )
\end{align}
 and 
 \begin{align}
	{\Pi}^{\pm}_{B_j}=(\frac{1 \pm \alpha \pm \eta}{2}P^{+}_{B_j}  + \frac{1 \pm \alpha \mp \eta}{2} P^{-}_{B_j} )
 \end{align}
 respectively. Following the similar approach adopted for unbiased POVMs in ontological model, local bound of CHSH inequality for biased POVMs is given can be calculated as
\begin{eqnarray}
\label{usb1}
\nonumber
\langle\mathbb{B}\rangle^{\alpha,\eta}_{\lambda} = \int _\Lambda \mu(\lambda|\rho) \mathbb{B}^{\alpha,\eta}_\lambda d\lambda \leq  2 ( |\alpha| +  \eta)^2
\\
\end{eqnarray}
The explicit derivation of Eq.(\ref{usb1}) is given in Appendix B. The local bound of CHSH inequality for biased POVM in Eq.(\ref{usb1}) is then dependent on both sharpness and biasedness parameters satisfying $\langle\mathbb{B}\rangle^{\alpha,\eta}_{\lambda} \leq  \langle\mathbb{B}\rangle^{S}_{\lambda}$.\\

 Now in QM, if Alice and Bob shares the same state given by Eq.(\ref{wer}) and same measurement directions used for earlier unbiased spin-POVM, the quantum bound of CHSH inequality is given by
\begin{eqnarray}
\label{beq}
\langle\mathbb{B}\rangle^{\alpha,\eta}_{Q} & \leq & \sqrt{2} \eta  (\alpha  \cos (2 \theta )+ \eta  \sin (2 \theta )+ \eta ) \\
\nonumber &+& 2 \alpha (\alpha + \eta \cos (2 \theta ))
 \end{eqnarray}
\begin{figure}[htp]
 \label{fig2}
 \centering
 \includegraphics[width=.35\textwidth]{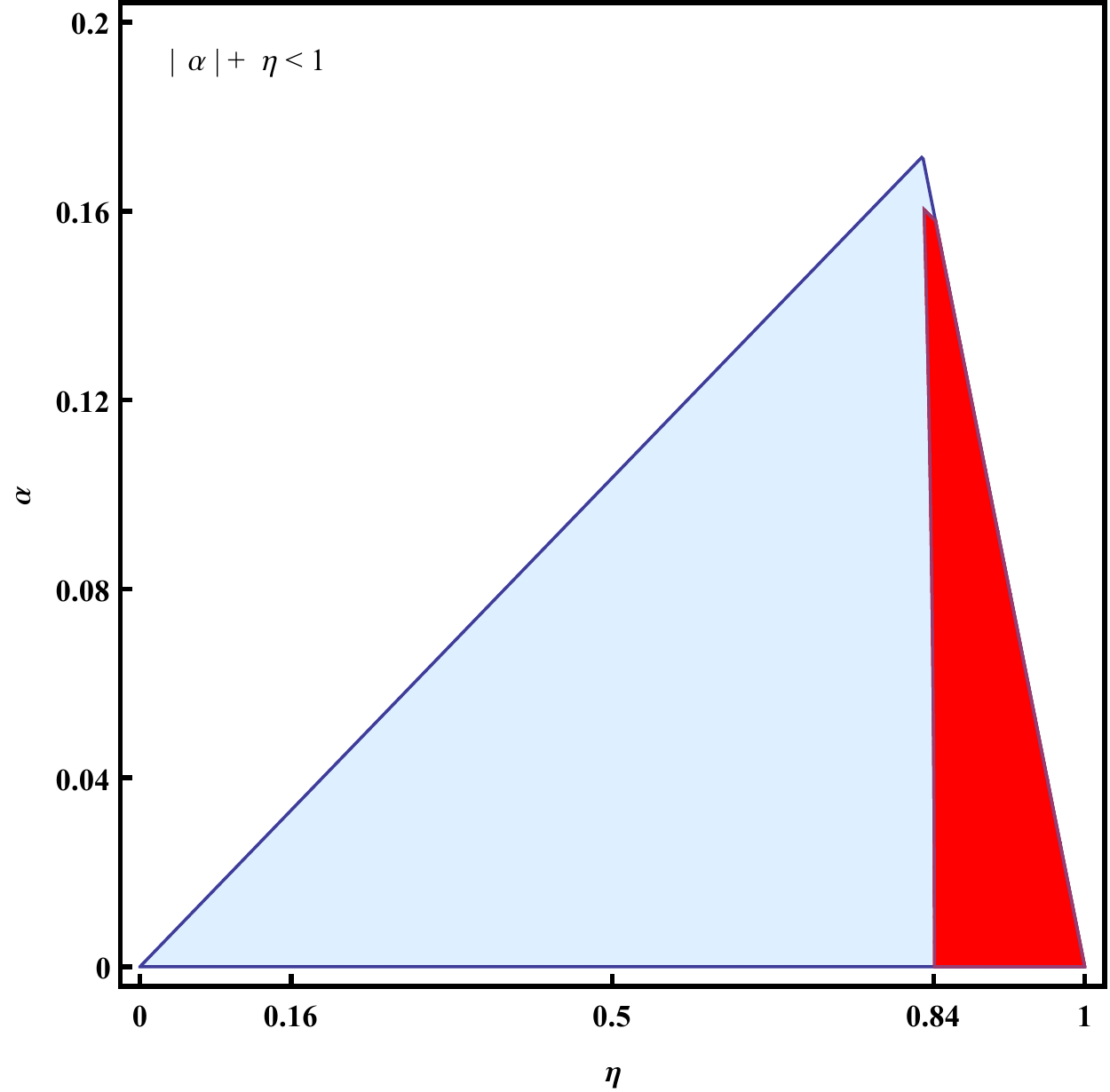}
 \caption{(color online) Shaded region formed by  $\Delta^{'S}>0$ (Red) and  $\Delta^{\alpha,\eta}>0$ (Blue or Red) shows the violation of standard CHSH inequality at  $\theta=\pi/4$  given $|\alpha|+\eta<1$.}
 \end{figure}
 The standard violation of CHSH inequality is obtained if 
 \begin{eqnarray}
\Delta^{'S}&=&\langle\mathbb{B}\rangle^{\alpha,\eta}_{Q}-\langle\mathbb{B}\rangle^{S}_{\lambda}\\ \nonumber
&=& \big[\sqrt{2} \eta  (\alpha  \cos (2 \theta )+ \eta  \sin (2 \theta )+\eta ) \\
\nonumber &+& 2 \alpha (\alpha + \eta \cos (2 \theta ))-2\big]>0
\end{eqnarray}
and in our revised scenario, we have
 \begin{eqnarray}
\Delta^{\alpha,\eta}&=&\langle\mathbb{B}\rangle^{\alpha,\eta}_{Q}-\langle\mathbb{B}\rangle^{\alpha, \eta}_{\lambda}\\ \nonumber
&=& \big[\sqrt{2} \eta  (\alpha  \cos (2 \theta )+ \eta  \sin (2 \theta )+ \eta ) \\
\nonumber &+& 2 \alpha (\alpha + \eta \cos (2 \theta ))- 2 ( |\alpha| +  \eta)^2\big] > 0
\end{eqnarray}
Since $\langle\mathbb{B}\rangle^{S}_{\lambda}\geq \langle\mathbb{B}\rangle^{\alpha,\eta}_{\lambda}$ then $\Delta^{\alpha,\eta}\geq \Delta^{'S}$. Here both $\Delta^{'S}>0$ and $\Delta^{\alpha,\eta}>0$ are dependent on sharpness and biasedness parameter. For $\alpha=0$,  $\Delta^{\alpha,\eta}$ becomes $\Delta^{\eta}$ given by Eq.($\ref{sdel}$) and hence the violation can be obtained whenever one has quantum violation of CHSH inequality for sharp measurements. In order to compare the violation of CHSH inequality in standard case and our modified scenario we plotted $\Delta^{\alpha,\eta}$ and $\Delta^{'S}$ for $\theta=\pi/4$. The shaded region in Figure.($1$) obtained for  $\Delta^{'S}>0$ (Red) and  $\Delta^{\alpha,\eta}>0$ (Blue and Red) at $\theta=\pi/4$.

 From the shaded region in Figure.($1$) we can say that the nonlocal region formed by the violation of standard CHSH inequality for biased POVMs is smaller than the violation of our revised local bound. It is found that violation of revised local bound of CHSH inequality (i.e., 
$\Delta^{\alpha,\eta}>0$) for biased POVMs is obtained for any value of $\eta$, for a range of $\alpha \in [0,0.16]$. But, in the standard case the violation is restricted by $0.84 \leq \eta \leq 1$ for the case when $\theta=\pi/4$.  Hence our revised bound on CHSH inequality for biased POVMs leads us to reveal nonlocality for a larger range of sharpness parameter than the standard one.     

\section{Summary and Discussion}
 We derive a modified local bound for Bell-CHSH inequality for both unbiased and biased POVMs. For the case of unbiased spin-POVMs, we have shown that for any given state, if there is violation for sharp measurement, there will be violation for unsharp measurement also. In the case of biased measurement and pure entangled state, we found the quantum violation of the modified CHSH inequality for a large range of sharpness and biasedness parameter when there will be no violation if one do not consider the modified local bound.  

\section*{Acknowledgments}
We gratefully acknowledge the insightful interaction with Prof. Guruprasad Kar (ISI Kolkata). AKP acknowledges the support from Ramanujan Fellowship research grant(SB/S2/RJN-083/2014).

\appendix
\begin{widetext}
\section{Local bound of CHSH inequality for spin-POVMs}
Noting that the hidden variable $\lambda$ only assign the response function to the projectors, the response fuctions for unbiased measurement in ontological model is convex combination of response function of projectors. Then, response function of unbiased measurement on Alice and Bob side are given by 
\begin{eqnarray}
\label{sa}
\xi^{\eta}(a_i|\Pi^{a_i}_{A_i},\lambda) = \frac{1 \pm \eta}{2} \xi(+1|P^{+}_{A_i},\lambda) + \frac{1 \mp \eta}{2} \xi(-1|P^{-}_{A_i},\lambda)
\end{eqnarray}
and
\begin{eqnarray}
\label{sb}
\xi^{\eta}(b_j|\Pi^{b_j}_{B_j},\lambda) = \frac{1 \pm \eta}{2} \xi(+1|P^{+}_{B_j},\lambda) + \frac{1\mp \eta}{2} \xi(-1|P^{-}_{B_j},\lambda)
\end{eqnarray}
respectively, where $\xi^{\eta}(a_i|\Pi^{a_i}_{A_i},\lambda), \xi^{\eta}(b_j|\Pi^{b_j}_{B_j},\lambda)  \in [0,1]$. 
Let us assume that in the case of unsharp measurement, CHSH expression in ontological model can be written as
\begin{eqnarray}
\label{usb}
\langle\mathbb{B}\rangle^{\eta}_{\lambda} =  \langle A_1 B_1\rangle^{\eta}_\lambda+\langle A_1 B_2\rangle^{\eta}_\lambda+\langle A_2 B_1\rangle^{\eta}_\lambda-\langle A_2 B_2\rangle^{\eta}_\lambda
\end{eqnarray}
where the joint expectation value of the observables, $\langle A_i B_j\rangle^{\eta}_{\lambda}$ followed by assumption of factorizability is given by
\begin{eqnarray}
\label{usa}
\langle A_i B_j\rangle^{\eta}_{\lambda}&=&\int_{\Lambda}\mu(\lambda|\rho)\sum_{a_i,b_j }a_i b_j\xi^{\eta}( a_i b_j|\Pi^{a_i}_{A_i},\Pi^{b_j}_{B_j},\lambda)  d\lambda \nonumber \\  &=& \int _\Lambda  \mu(\lambda|\rho) \overline{p}( a_i |\Pi^{a_i}_{A_i},\lambda) \overline{p}( b_j|\Pi^{b_j}_{B_j},\lambda)  d\lambda
\end{eqnarray}

Here $\overline{p}( a_i |\Pi^{a_i}_{A_i},\lambda)=\sum_{a_i} a_i \xi^{\eta}( a_i|\Pi^{a_i}_{A_i},\lambda) $ and $\overline{p}( b_j|\Pi^{b_j}_{B_j},\lambda)=\sum_{b_i} b_i \xi^{\eta}( b_i|\Pi^{b_j}_{B_j},\lambda)$

Using Eq.(\ref{sa}) and Eq.(\ref{sb}) we obtain
\begin{eqnarray}
\label{usa1b1}
  \langle A_1 B_1\rangle^{\eta}_\lambda&=&\eta^2 \big(\xi(+1|P^{+}_{A_1},\lambda )-\xi(-1|P^{-}_{A_1},\lambda)\big) \big(\xi(+1|P^{+}_{B_1},\lambda)-\xi(-1|P^{-}_{B_1},\lambda)\big)
\end{eqnarray}

Similarly
\begin{eqnarray}
\label{usa1b2}
  \langle A_1 B_2\rangle^{\eta}_\lambda &=& \eta^2 \big(\xi(+1|P^{+}_{A_1},\lambda )-\xi(-1|P^{-}_{A_1},\lambda)\big) \big(\xi(+1|P^{+}_{B_2},\lambda)-\xi(-1|P^{-}_{B_2},\lambda)\big)
\end{eqnarray}

\begin{eqnarray}
\label{usa2b1}
 \langle A_2 B_1\rangle^{\eta}_\lambda&=&\eta^2 \big(\xi(+1|P^{+}_{A_2},\lambda )-\xi(-1|P^{-}_{A_2},\lambda)\big) \big(\xi(+1|P^{+}_{B_1},\lambda)-\xi(-1|P^{-}_{B_1},\lambda)\big)
\end{eqnarray}

and

\begin{eqnarray}
\label{usa2b2}
  \langle A_2 B_2\rangle^{\eta}_\lambda&=& \eta^2 \big(\xi(+1|P^{+}_{A_2},\lambda )-\xi(-1|P^{-}_{A_2},\lambda)\big) \big(\xi(+1|P^{+}_{B_2},\lambda)-\xi(-1|P^{-}_{B_2},\lambda)\big)
\end{eqnarray}
Putting Eq.($\ref{usa}-\ref{usa2b2}$) in  Eq.($\ref{usb}$) we obtain the local bound of CHSH inequality for unbiased POVMs is given below
\begin{eqnarray}
\label{bell}
\langle\mathbb{B}\rangle^{\eta}_{\lambda} \leq 2\eta^2
\end{eqnarray}

\section{Local bound of CHSH inequality for biased POVMs} 
Similarly to the case of unbiased POVMs, the response functions of biased POVMs can be written as
\begin{eqnarray}
\label{a1p}
\xi^{\alpha,\eta}(a_i|\Pi^{a_i}_{A_i},\lambda) = \frac{1\pm\alpha\pm\eta}{2} \xi(+1|P^{+}_{A_i},\lambda ) + \frac{1\pm\alpha\mp\eta}{2} \xi(-1|P^{-}_{A_i},\lambda )
\end{eqnarray}
\begin{eqnarray}
\label{b1p}
\xi^{\alpha,\eta}(b_j|\Pi^{b_j}_{B_j},\lambda) = \frac{1\pm\alpha\pm\eta}{2} \xi(+1|P^{+}_{B_j},\lambda ) + \frac{1\pm\alpha\mp\eta}{2}\xi(-1|P^{-}_{B_j},\lambda )
\end{eqnarray}
respectively, where $\xi^{\alpha,\eta}(a_i|\Pi^{a_i}_{A_i},\lambda),\xi^{\alpha,\eta}(b_j|\Pi^{b_j}_{B_j},\lambda)  \in [0,1]$. \\

We can then write 
\begin{eqnarray}
\label{ba1b1}
\nonumber
\langle A_1 B_1\rangle^{\alpha,\eta}_\lambda &=& \bigg((\alpha+\eta)\xi(+1|P^{+}_{A_1},\lambda )-(\eta-\alpha)\xi(-1|P^{-}_{A_1},\lambda)\bigg) \bigg((\alpha+\eta)\xi(+1|P^{+}_{B_1},\lambda)-(\eta-\alpha)\xi(-1|P^{-}_{B_1},\lambda)\bigg)\\ 
 &=&  \eta^2 \langle A_1 B_1\rangle_\lambda + \alpha^2 + 2\eta \alpha \bigg(\xi(+1|P^{+}_{A_1},\lambda )\xi(+1|P^{+}_{B_1},\lambda)-\xi(-1|P^{-}_{A_1},\lambda )\xi(-1|P^{-}_{B_1},\lambda)\bigg),
\end{eqnarray}
Similarly
\begin{eqnarray}
\label{ba1b2}
\nonumber
 \langle A_1 B_2\rangle^{\alpha,\eta}_\lambda &=&  \eta^2 \langle A_1 B_2\rangle_\lambda + \alpha^2 +2 \eta \alpha \bigg(\xi(+1|P^{+}_{A_1},\lambda )\xi(+1|P^{+}_{B_2},\lambda )-\xi(-1|P^{-}_{A_1},\lambda )\xi(-1|P^{-}_{B_2},\lambda)\bigg),
 \\
\end{eqnarray}

\begin{eqnarray}
\label{ba2b1}
\nonumber
\langle A_2 B_1\rangle^{\alpha,\eta}_\lambda&=&  \eta^2 \langle A_2 B_1\rangle_\lambda + \alpha^2 +2 \eta \alpha \bigg(\xi(+1|P^{+}_{A_2},\lambda )\xi(+1|P^{+}_{B_1},\lambda )-\xi(-1|P^{-}_{A_2},\lambda )\xi(-1|P^{-}_{B_1},\lambda)\bigg),
\\
\end{eqnarray}

and
\begin{eqnarray}
\label{ba2b2}
\nonumber
 \langle A_2 B_2\rangle^{\alpha,\eta}_\lambda&=& \eta^2 \langle A_2 B_2\rangle_\lambda + \alpha^2 +2 \eta \alpha \bigg(\xi(+1|P^{+}_{A_2},\lambda )\xi(+1|P^{+}_{B_2},\lambda )-\xi(-1|P^{-}_{A_2},\lambda )\xi(-1|P^{-}_{B_2},\lambda)\bigg).
 \\
\end{eqnarray}

Using Eqs.($\ref{ba1b1}-\ref{ba2b2}$), the CHSH expression in ontological model for biased measurement is obtained as

\begin{eqnarray}
\label{biaschsh}
\langle\mathbb{B}\rangle^{\alpha,\eta}_{\lambda} = \eta^2 \langle\mathbb{B}\rangle^S_{\lambda} + 2 \alpha^2 + 4 \eta \alpha \bigg(\xi(+1|P^{+}_{B_1},\lambda )-\xi(-1|P^{-}_{A_1},\lambda )\bigg) 
\end{eqnarray}
where $\langle\mathbb{B}\rangle^S_{\lambda}$ is the CHSH expression for projectors. Now, maximizing  Eq.($\ref{biaschsh}$), the CHSH inequality for biased POVMs can be written as
\begin{eqnarray}
\langle\mathbb{B}\rangle^{\alpha,\eta}_{\lambda}& \leq & 2 ( | \alpha| +  \eta)^2
\end{eqnarray}
which is  Eq.(\ref{usb1}) in the main text.

\end{widetext}

\end{document}